\documentclass[11pt]{article}

\usepackage[final]{acl}


\usepackage{booktabs}
\usepackage{multirow}
\usepackage{times}
\usepackage{latexsym}

\usepackage[T1]{fontenc}

\usepackage[utf8]{inputenc}

\usepackage{microtype}

\usepackage{inconsolata}

\usepackage{graphicx}

\usepackage{lipsum}

\usepackage[table]{xcolor}
\definecolor{g1}{HTML}{E3F2FD}
\definecolor{g2}{HTML}{E8F5E9}
\definecolor{g3}{HTML}{FFF8E1}
\definecolor{g4}{HTML}{FCE4EC}
\definecolor{g5}{HTML}{F3E5F5}

\title{From Metrics to Natural Dialogue: French Full-Duplex Benchmark for Spoken Dialogue Models}

\author{
  Hamid Soltani \and Gilles Boulianne \\
  Luqia Technologies, Montréal, Québec, Canada \\
  \texttt{hamid.soltani@concordia.ca \hfill gilles.boulianne@luqia.ca} \\
}

\begin{document}
\maketitle
\begin{abstract}
Full-duplex spoken dialogue models aim to make voice agents more natural by allowing them to listen, speak, pause, and respond during ongoing conversation. However, it is not clear whether full-duplex benchmarks behave the same way when models are evaluated in a different language. To investigate this, we introduce a French full-duplex benchmark (FDB) with two variants, CALLFC-FDB for Canadian French and MEDIA-FDB for European French, and compare them with an English FDB. Built from real spoken resources, these benchmarks evaluate key full-duplex skills, including pause handling, turn-taking, backchannels, and interruptions. Beyond introducing French FDBs, we evaluate human--human conversations with FDB metrics to better understand the values these metrics take in real-world dialogue. Our analysis reveals that most timing-based metrics behave similarly across languages, while content-based evaluation degrades under language mismatch. We also find a trade-off between optimizing benchmark metrics and preserving conversational naturalness.

\end{abstract}


\section{Introduction}

Human-computer interaction has long studied how machines can move beyond command execution toward understanding human intention through natural communication. This goal aligns with agentic AI, where systems perceive context, make decisions, and act toward user goals \citep{xi_rise_2023}. Speech is one of the most natural forms of this interaction, enabling direct communication between users and AI systems \citep{ji2024}. Unlike text, speech carries words together with timing, hesitation, emphasis, and other acoustic cues. These signals make speech a rich form of interaction, but they also make spoken dialogue difficult, since systems must understand language, interpret acoustics, handle pauses and overlap, and respond in real time. Spoken Dialogue Models (SDMs) \citep{defossez2024a, roy2026} address this challenge by taking speech as input and generating spoken responses. A key distinction among SDMs is how they manage conversational timing through listening, speaking, pausing, and yielding. Based on this, SDMs are divided into half-duplex and full-duplex systems \citep{lin2025g}. In half-duplex interaction, the model responds after detecting the end of the user turn. This simplifies processing, but differs from human conversation, where speakers coordinate turns through pauses, overlap, and listener feedback \citep{heldner2010}. Full-duplex SDMs reduce this gap by listening and speaking at the same time, supporting turn-taking \citep{raux2012}, backchannels \citep{schegloff1982discourse}, and overlapping speech \citep{schegloff2000}.

Full-duplex SDMs generally follow two architectural families: cascaded and speech-to-speech systems. In cascaded systems, speech-to-text (STT) converts user speech into text, a large language model (LLM) generates a response, and text-to-speech (TTS) converts it back into speech. AudioGPT \citep{huang2023e} connects ChatGPT with automatic speech recognition (ASR), TTS, and audio foundation models, keeping the LLM as the reasoning center while separate audio models handle speech recognition, synthesis, understanding, and generation. Recently, DuplexCascade \citep{yang2026} uses streaming ASR micro-turns and control tokens to decide whether to wait, respond, or backchannel. Overall, cascaded systems remain attractive because they reuse strong speech and language components and allow each part to be improved independently. However, they can add latency and lose acoustic information during STT and TTS conversion.

In contrast, speech-to-speech systems process and generate speech directly. This makes them more suitable for full-duplex interaction, where models must listen while speaking and react to interruptions, backchannels, and overlap. Freeze-Omni \citep{wang2024e} moves closer to real-time spoken dialogue by connecting a streaming speech encoder and speech decoder to a frozen text LLM, enabling low-latency speech-to-speech interaction. Moshi \citep{defossez2024a} represents user and system speech as parallel streams, allowing it to handle overlaps and interruptions without fixed speaker turns. PersonaPlex \citep{roy2026} extends this direction with role conditioning and voice control, using hybrid prompts that combine text-based role instructions with speech samples for voice cloning. 
Together, these models show that speech-to-speech systems preserve acoustic and conversational information needed for natural full-duplex interaction.

As full-duplex SDMs become more realistic, evaluation becomes more difficult. Recent work has introduced interaction-focused speech datasets that capture behaviors such as interruptions and backchannels \citep{chen-etal-2025-interactspeech}. Beyond answer quality, evaluation must measure whether a model responds at the right time, stops when needed, ignores irrelevant speech, and remains consistent across turns. Full-Duplex-Bench (FDB) v1.0 \citep{lin2025g} evaluates the interaction skills required for full-duplex dialogue, including pause handling, backchanneling, smooth turn-taking, and user interruption. The benchmark focuses on turn-taking behavior, assessing when the model should wait, backchannel, or take the turn. FDB v1.5 \citep{lin2025f} extends this setting to overlapping speech by separating user interruptions, user backchannels, speech directed to others, and background speech, since each case requires a different system behavior. FDB v2 \citep{lin2025e} moves from single-event testing to multi-turn dialogue. It uses an automated spoken examiner with staged goals, follow-up questions, interruptions, and pacing styles. This allows evaluation of turn-taking fluency, multi-turn instruction following, and task-specific performance across daily assistance, correction handling, entity tracking, and safety.

One important open question is whether the test language itself affects the evaluation of full-duplex spoken dialogue models. In particular, it is unclear how FDB metrics behave when a target-language model is evaluated using a non-target language benchmark. This question matters because full-duplex evaluation measures not only response quality, but also timing-sensitive behaviors such as pauses, backchannels, turn-taking, and interruptions. Although recent multilingual benchmarks have expanded evaluation beyond English \citep{ma-etal-2025-c3}, dedicated benchmarks for French full-duplex dialogue are still missing. French also contains language-specific interactional cues such as \textit{euh}, \textit{ben}, \textit{ouais}, and \textit{d'accord}, which do not map directly onto English translations. Their meanings and conversational functions depend on context, prosody, and turn-taking. In addition, some existing FDB tasks are based on synthetic speech, which may not fully reflect the variability of natural conversations. To study the effect of evaluation language and to support research on French spoken dialogue systems, we introduce two real-world French Full-Duplex-Bench variants: CALLFC-FDB for Canadian French and MEDIA-FDB for European French.\footnote{Dataset preparation code: \url{https://github.com/crim-ca/fdb-eval-prep-CCCF}} CALLFC-FDB is built from the CALLFRIEND French Canadian \citep{mondada2006cabank, canavan1996callfriend}, a corpus of telephone conversations between native speakers of Canadian French, while MEDIA-FDB is built from the MEDIA \citep{bonneau-maynard_semantic_2005}, a French spoken language understanding corpus. Together, they support evaluation of French voice agents across different French varieties and interaction settings. Our results demonstrate that most timing-based metrics remain broadly consistent across languages, while content-based evaluation is more sensitive to language mismatch.

Beyond extending FDBs to French, we evaluate human-human conversations with FDB metrics to establish reference values for how these metrics behave in real dialogue. Unlike agent-generated interactions, these conversations provide a natural reference point for interpreting metric-based evaluation. This is important since current FDB metrics define better performance through metric optimization. However, real dialogue is more flexible than fixed metric targets, and our analysis shows a trade-off between optimizing FDB metrics and maintaining natural spoken interaction.

\section{French Full-Duplex-Bench}

\subsection{FDB Tasks}

To ensure compatibility with the FDB series, we adapt the four core tasks from FDB v1.0 \citep{lin2025g}. We focus on FDB v1.0 as a first step because it defines the foundational full-duplex skills required for spoken dialogue evaluation.

\begin{itemize}
    \item \textbf{Pause Handling}: whether the model remains silent when the user pauses inside a turn.
    
    \item \textbf{Backchanneling}: whether the model produces appropriate backchannels without taking the turn.

    \item \textbf{Smooth Turn Taking}: whether the model responds after the user finishes speaking, without undue latency.

    \item \textbf{User Interruption}: whether the model can handle interruptions and shift from the previous direction to the new intent.
\end{itemize}

For MEDIA-FDB, we did not find enough valid examples of system backchannels, so the backchanneling task is not included for this dataset. These tasks may depend on the evaluation language in different ways. Pause handling is expected to be weakly language-dependent, since the model mainly needs to detect silence inside an ongoing turn. Smooth turn-taking also relies on timing, but may be more affected by language matching because the model must decide whether the user has completed an utterance and then take the turn. Backchanneling is more language-dependent, because the model must decide when short listener feedback is needed and produce it with appropriate timing and duration. User interruption is the most language-dependent task, as the model must understand the interruption and respond to the new intent.

\subsection{FDB Metrics}

We briefly review the FDB v1.0 \citep{lin2025g} metrics used in our evaluation:

\begin{itemize}
\item \textbf{Takeover Rate (TOR):} average takeover decision across samples, where silence and backchannels are scored as 0, and other responses as 1. Lower TOR is preferred for pause handling and backchanneling, while higher TOR is better for smooth turn-taking and user interruption.

\item \textbf{Backchannel Frequency:} number of backchannels normalized by duration. Higher values show more frequent listener feedback.

\item \textbf{Jensen-Shannon Divergence (JSD):} measures the difference between model and ground-truth backchannel timing distributions. Lower values indicate greater similarity between model and ground-truth backchannels locations and durations.

\item \textbf{Smooth Turn-Taking Latency:} time between the end of the user turn and the start of the model response.

\item \textbf{User Interruption Latency:} time between a user interruption and the model response.

\item \textbf{Rating:} LLM-based response-quality score for user interruption, from 0 to 5.
\end{itemize}

Most FDB metrics are based on voice activity, such as the start of speech, silence, or the presence of backchannels. They do not directly evaluate speech content, so they are basically language-independent. The main exception is the rating metric for user interruption, which depends on response content. For example, an English response to a French interruption would receive a low rating even if the timing behavior is appropriate. Latency can also be influenced by system architecture and implementation, especially in cascaded pipelines.

\subsection{CALLFRIEND and MEDIA Data}

We build the French Full-Duplex-Bench from two French spoken-dialogue corpora:
CALLFRIEND French Canadian and MEDIA.

\subsubsection{CALLFRIEND French Canadian}

CALLFRIEND French Canadian contains 26 hours of unscripted telephone conversations between native speakers of Canadian French, recorded in 1996 and distributed by the Linguistic Data Consortium \citep{canavan1996callfriend}. In each conversation, participants could call a person of their choice and talk about any topic. Most speakers called family or friends, resulting in highly informal speech.

\subsubsection{MEDIA}

MEDIA is a French human-machine dialogue corpus in the hotel reservation and tourist information domain. It was collected using a
Wizard-of-Oz (WoZ) setup \citep{green_rapid_1985}, where users believed they were interacting with an
intelligent system, while a human operator produced the system responses
\citep{laperriere2022b}. Unlike CALLFRIEND, MEDIA contains task-oriented conversations with clearer interaction goals.

Together, CALLFRIEND and MEDIA increase benchmark diversity by covering informal open-domain and structured task-oriented dialogue.

\subsection{French FDB Dataset Preparation}

We construct the French Full-Duplex-Bench from continuous two-channel spoken
conversations. The preparation process converts natural French speech into FDB-compatible samples, while preserving the timing, speaker activity, pauses, and backchannels that are important for full-duplex evaluation. CALLFC-FDB and MEDIA-FDB share the same output format, scoring, materialization, and export stages. However, they differ in transcript preparation and turn extraction.

\subsubsection{Transcription Process}
Because the two corpora differ in their transcription
annotations, we use two corpus-specific transcript preparation methods.

\paragraph{CALLFC-FDB transcript preparation.}
The CALLFRIEND FC conversations were manually transcribed by in-house
annotators with speaker turn time boundaries, but without speaker assignment. Since overlapping regions include only the dominant speaker, we apply additional processing to recover speaker information from these regions.

To recover as much speaker information as possible from overlapped regions, we first transcribe each audio channel with Whisper large-v3 \citep{radford}, and align the outputs with the Montreal Forced Aligner (MFA) \citep{mcauliffe2017}. We align the automatic and manual transcripts using Longest Common Subsequence (LCS) \citep{hirschberg_algorithms_1977}, allowing us to transfer word-level timings and speaker assignments from the automatic transcripts to the manual transcription. The final merged CALLFC-FDB transcript keeps all words from the manual transcription and adds automatic transcription words in overlapped regions that were not present in the manual transcript. This step recovers 44,165 overlapped words, about 11\% of the final transcript of 386K words. The result is a word-level, speaker-assigned transcript that combines the reliable words from the manual transcription with timing and channel information from the automatic channel-level alignment.

\paragraph{MEDIA-FDB transcript preparation.}
The manual MEDIA transcriptions already include speaker assignments and turn-level overlap annotations. The manual transcriptions for each channel are directly aligned with the corresponding audio using the MFA, obtaining word-level timestamps while keeping the manually transcribed words. Thus, MEDIA-FDB
is based on manually transcribed text with MFA-derived word-level timing.

\subsubsection{Turn and sentence extraction}
FDB tasks consist of one audio file, \texttt{input.wav}, containing an extract of conversation corresponding to the task, and an optional JSON annotation file required for some evaluation metrics. FDB tasks are extracted from pairs of speaker turns, with one turn coming from
each speaker. The definition of a turn depends on the corpus. For MEDIA-FDB, speaker turns are taken directly from the manual transcription. For CALLFC-FDB, where the manual transcription does not provide the same role structure, words within chunks separated by no more than 2.0 seconds are considered as part of the same turn. After turns are obtained, each turn is split into sentence-like units using timing gaps between words. Specifically, a new sentence is created when the inter-word gap is greater than 0.2 seconds. Thus, a turn may contain one or several sentences. Since CALLFC-FDB contains two human speakers with no fixed user or assistant roles, we process it twice. In the first pass, one speaker is treated as the speaker side and the other as the WoZ side. In the second pass, the roles are reversed.

\subsubsection{FDB event extraction}

We select candidate samples from paired speaker turns, closely following \cite{lin2025g}:

\begin{itemize}
    \item \textbf{Pause handling.}
    This task operates on the full speaker turn, including all of its sentences. The pipeline detects inter-word gaps longer than 0.7 seconds within
    the turn. Each gap becomes a separate candidate sample. 

    \item \textbf{Backchanneling.}
    For speaker turns of at least 2.0 seconds, we identify isolated backchannel words from the WoZ side speaker. 

    \item \textbf{Smooth turn-taking.}
    We select instances where the gap between the speaker turn and the following WoZ turn is less than 1.0s. We use the last sentence of the speaker with 5 seconds of silence appended. 

    \item \textbf{User interruption.}
    We identify instances where the speaker starts talking before the previous WoZ turn has finished and the speaker sentence is not a backchannel. 
\end{itemize}

Samples that satisfy the necessary and sufficient classification conditions for
a task are first selected as candidates. A single turn pair may produce more
than one candidate when it contains multiple valid events, for example multiple
qualifying pauses.

\subsubsection{Scoring, validation, and export}
Candidate selection is followed by a scoring and validation stage.
This stage applies hard post-hoc requirements to remove samples that passed the
initial classification step but are still unsuitable as evaluation examples.
These checks include duration limits, logical consistency checks, and
task-specific timing constraints. For example, valid pauses are positive and fall inside the turn, backchannel inputs last at least 15 seconds, and user interruption contexts should be between 0.5 and 5.0 seconds. After hard filtering, 92.8\% of MEDIA candidates and 78.0\% of CALLFC candidates were retained.

The scoring process also assigns a quality score to each candidate. This score
measures how clearly and unambiguously the sample represents the intended task.
As examples, for \texttt{pause\_handling}, the ideal pause is around 0.7
seconds, with enough speech before and after it. For
\texttt{smooth\_turn\_taking}, shorter inter-turn gaps receive higher scores. Finally, only validated candidates that pass all filters are used to create the
final FDB samples. For each selected event, the pipeline extracts the required
audio segment from the original recording. The highest-quality validated candidates
are then materialized and exported in the standard FDB directory format as
MEDIA-FDB and CALLFC-FDB. To facilitate comparison with the original English FDB benchmark, the number of samples per task was capped at 150 whenever sufficient candidates were available. After quality-based ranking and the sample cap, the final datasets contained 41.4\% and 26.3\% of the original MEDIA and CALLFC candidates, respectively. Table~\ref{tab:french-fdb-counts} reports the number
of extracted samples for each task in both datasets.

\section{Evaluation}
We evaluate the proposed French FDBs using both English and French SDMs. For the English setting, we use Freeze-Omni \citep{wang2024e}\footnote{\url{https://github.com/VITA-MLLM/Freeze-Omni}} and PersonaPlex-7B-v1 \citep{roy2026}\footnote{\url{https://github.com/NVIDIA/personaplex}} as reference models to compare English FDB with the proposed French FDBs. Freeze-Omni was evaluated on an NVIDIA A40 48GB GPU, while PersonaPlex was evaluated on an NVIDIA A100 80GB GPU.

For a fair English/French comparison, we evaluate two cascaded SDMs based on an STT--LLM--TTS pipeline and implemented with Pipecat\footnote{\url{https://github.com/pipecat-ai/pipecat}}. Vosk is used for STT, with \texttt{vosk-model-small-en-us-0.15} for English and \texttt{vosk-model-small-fr-0.22} for French\footnote{\url{https://alphacephei.com/vosk/models}}. Both systems use \texttt{qwen3:8b} through Ollama\footnote{\url{https://ollama.com/library/qwen3:8b}} and Piper TTS, using \texttt{en\_US-lessac-low} for English and \texttt{fr\_FR-gilles-low} for French\footnote{\url{https://github.com/rhasspy/piper}}. All inference experiments for the cascaded systems were conducted on an NVIDIA A40 GPU with 48 GB of memory. Their relatively high latency is expected from the simple cascaded design and is not specific to the French FDB datasets, since similar latency is also observed on English FDB. Our logs show that STT and TTS latency are small, while most delay comes from LLM generation. To assess STT quality, we measured Vosk WER on the pause-handling subsets, which contain longer user speech segments. For example, the WER is 21.7\% for French on MEDIA-FDB and 26.7\% for English on English FDB. We do not report mismatched-language WER because the recognizers are language-specific and the WER is expected to be very high. We also include human--human baselines for both French FDB datasets as a natural reference for interpreting the metrics.

\begin{table}[t]
\centering
\small
\setlength{\tabcolsep}{6pt}
\begin{tabular}{@{}p{0.50\columnwidth}cc@{}}
\toprule
\multirow{2}{*}{\textbf{Task}} &
\multicolumn{2}{c}{\hspace{0.7em}\textbf{Number of Samples}} \\
\cmidrule(l){2-3}
& \textbf{CALLFC} & \textbf{MEDIA} \\
\midrule
Pause handling & 150 & 150 \\
Backchanneling & 37 & -- \\
Smooth turn-taking & 150 & 96 \\
User interruption & 150 & 38 \\
\bottomrule
\end{tabular}
\caption{Number of extracted French FDB samples for each task in CALLFC-FDB and MEDIA-FDB. A dash indicates that no valid samples were available.}
\label{tab:french-fdb-counts}
\end{table}

\begin{table*}[t]
\centering
\small
\setlength{\tabcolsep}{4pt}
\resizebox{\textwidth}{!}{
\begin{tabular}{llccccccccc}
\toprule
\textbf{Data} & \textbf{Model} & \multicolumn{1}{c}{\textbf{Pause Handling}} & \multicolumn{3}{c}{\textbf{Backchannel}} & \multicolumn{2}{c}{\textbf{Smooth Turn Taking}} & \multicolumn{3}{c}{\textbf{User Interruption}} \\
\cmidrule(lr){3-3} \cmidrule(lr){4-6} \cmidrule(lr){7-8} \cmidrule(lr){9-11}
& & TOR $\downarrow$ & TOR $\downarrow$ & Freq $\uparrow$ & JSD $\downarrow$ & TOR $\uparrow$ & Latency $\downarrow$ & TOR $\uparrow$ & Rating $\uparrow$ & Latency $\downarrow$ \\
\midrule

\multirow{4}{*}{English FDB}
& Freeze-Omni (En) & \cellcolor{g3}0.530 & \cellcolor{g3}0.654 & \cellcolor{g2}0.031 & \cellcolor{g2}0.887 & \cellcolor{g3}0.319 & \cellcolor{g2}0.890 & \cellcolor{g2}0.565 & \cellcolor{g2}3.354 & \cellcolor{g2}1.068 \\
& PersonaPlex (En) & \cellcolor{g2}0.300 & \cellcolor{g2}0.255 & \cellcolor{g1}0.096 & \cellcolor{g1}0.756 & \cellcolor{g1}0.941 & \cellcolor{g1}0.322 & \cellcolor{g1}0.955 & \cellcolor{g1}4.702 & \cellcolor{g1}0.147 \\
& Cascade (En) & \cellcolor{g1}0.009 & \cellcolor{g1}0.091 & \cellcolor{g3}0.002 & \cellcolor{g3}0.989 & \cellcolor{g2}0.529 & \cellcolor{g3}7.157 & \cellcolor{g3}0.380 & \cellcolor{g2}3.211 & \cellcolor{g3}2.612 \\
& Cascade (Fr) & \cellcolor{g1}0.006 & \cellcolor{g2}0.182 & \cellcolor{g3}0.006 & \cellcolor{g3}0.960 & \cellcolor{g4}0.193 & \cellcolor{g3}7.861 & \cellcolor{g4}0.215 & \cellcolor{g3}0.070 & \cellcolor{g4}4.705 \\

\midrule
\multirow{5}{*}{CALLFC-FDB}
& Freeze-Omni (En) & \cellcolor{g3}0.540 & \cellcolor{g1}0.243 & \cellcolor{g3}0.015 & \cellcolor{g4}0.948 & \cellcolor{g2}0.360 & \cellcolor{g3}1.100 & \cellcolor{g3}0.593 & \cellcolor{g4}0.348 & \cellcolor{g2}1.107 \\
& PersonaPlex (En) & \cellcolor{g3}0.447 & \cellcolor{g3}0.486 & \cellcolor{g1}0.110 & \cellcolor{g2}0.818 & \cellcolor{g1}0.787 & \cellcolor{g2}0.883 & \cellcolor{g2}0.733 & \cellcolor{g4}0.145 & \cellcolor{g1}0.347 \\
& Cascade (En) & \cellcolor{g1}0.213 & \cellcolor{g3}0.649 & \cellcolor{g3}0.007 & \cellcolor{g4}0.970 & \cellcolor{g3}0.147 & \cellcolor{g4}7.816 & \cellcolor{g4}0.400 & \cellcolor{g3}0.833 & \cellcolor{g3}5.300 \\
& Cascade (Fr) & \cellcolor{g1}0.127 & \cellcolor{g3}0.486 & \cellcolor{g3}0.019 & \cellcolor{g3}0.912 & \cellcolor{g2}0.307 & \cellcolor{g4}8.252 & \cellcolor{g1}0.953 & \cellcolor{g2}1.385 & \cellcolor{g3}5.230 \\
& Human-Human (Fr) & \cellcolor{g1}0.193 & \cellcolor{g1}0.270 & \cellcolor{g1}0.114 & \cellcolor{g1}0.586 & \cellcolor{g1}0.773 & \cellcolor{g1}0.172 & \cellcolor{g1}0.860 & \cellcolor{g1}1.930 & \cellcolor{g2}1.585 \\

\midrule
\multirow{5}{*}{MEDIA-FDB}
& Freeze-Omni (En) & \cellcolor{g3}0.540 & -- & -- & -- & \cellcolor{g3}0.406 & \cellcolor{g2}1.041 & \cellcolor{g3}0.632 & \cellcolor{g3}0.417 & \cellcolor{g1}1.023 \\
& PersonaPlex (En) & \cellcolor{g2}0.227 & -- & -- & -- & \cellcolor{g2}0.802 & \cellcolor{g2}1.132 & \cellcolor{g1}0.868 & \cellcolor{g3}0.151 & \cellcolor{g1}0.596 \\
& Cascade (En) & \cellcolor{g1}0.060 & -- & -- & -- & \cellcolor{g3}0.500 & \cellcolor{g3}8.117 & \cellcolor{g1}0.974 & \cellcolor{g3}0.486 & \cellcolor{g3}4.069 \\
& Cascade (Fr) & \cellcolor{g2}0.180 & -- & -- & -- & \cellcolor{g3}0.438 & \cellcolor{g3}8.180 & \cellcolor{g1}0.895 & \cellcolor{g1}3.471 & \cellcolor{g3}4.572 \\
& Human-Human (Fr) & \cellcolor{g1}0.040 & -- & -- & -- & \cellcolor{g1}0.990 & \cellcolor{g1}0.329 & \cellcolor{g1}0.921 & \cellcolor{g1}2.829 & \cellcolor{g2}2.153 \\

\bottomrule
\end{tabular}
}
\caption{Evaluation of English and French spoken dialogue models on the original English FDB and the proposed French FDB datasets. Human--human conversations are included as a reference for natural conversational behavior. Within dataset and column, values have the same color if their 95\% confidence intervals overlap by more than 25\% (i.e. difference not statistically significant). English pause-handling values combine the two task IDs using pooled summary statistics.}
\label{tab:fdb_tasks_EN_FR}
\end{table*}

In the FDB framework, we modified the ASR step implemented in \texttt{asr.py} by replacing NVIDIA Parakeet-TDT-0.6b-v2 with the v3 version\footnote{\url{https://huggingface.co/nvidia/parakeet-tdt-0.6b-v3}}. This change was necessary because v2 does not support French, whereas v3 supports multilingual ASR and can generate the required \texttt{output.json} transcription files. This modification made the evaluation pipeline suitable for MEDIA-FDB and CALLFC-FDB. For the rating metric in the user-interruption task, we use \texttt{llama3.3:70b}\footnote{\url{https://ollama.com/library/llama3.3:70b}} as the LLM judge. We selected this large-scale LLM because it has a different architecture from the spoken dialogue models under evaluation, reducing the risk that the judge shares the same modeling biases as the evaluated systems. For the user-interruption rating, we follow the official FDB evaluation procedure, including the same prompt, scoring rubric, and generation settings.

\section{Results}

Our main goal is to study whether evaluation language affects FDB metrics. Table~\ref{tab:fdb_tasks_EN_FR} compares English and French models on the English and French FDB datasets, with human--human conversations as a reference. Full 95\% confidence intervals for all results are reported in Appendix~\ref{sec:confidence_intervals}.

As a sanity check, Freeze-Omni and PersonaPlex results on the original English FDB are generally consistent with FDB v1.0, giving us confidence in our setup. For pause handling, English FDB generally gives lower or comparable TOR values across both English and French models. This suggests that pause handling mainly depends on voice activity, such as silence inside a turn, rather than on the language of the evaluated model. For backchanneling, PersonaPlex gives the highest backchannel frequency and the lowest JSD among the evaluated models on both English FDB and CALLFC-FDB. This suggests that backchannel behavior is affected by model architecture, not only by the evaluation language or model language. For smooth turn-taking, cascaded models achieve higher TOR when the benchmark language matches the model language. Freeze-Omni remains stable, while PersonaPlex drops on the French benchmarks. This suggests some benefit from language matching for this task. For smooth turn-taking and user interruption, the English cascade consistently has lower latency than the French cascade across all FDB datasets. Since both systems use the same pipeline structure, this suggests that latency is shaped by system implementation and architecture, not only by the benchmark language. As expected, user-interruption rating is the most language-dependent metric. Ratings drop sharply when the model language does not match the benchmark language, while language-matched settings obtain much higher scores. When models are evaluated on non-target language datasets, most metrics based on voice activity remain broadly comparable. The main exception is content-based evaluation, such as user-interruption rating. Overall, language mismatch does not fundamentally change most timing-based FDB metrics, but it strongly affects content-based evaluation.

Human--human baselines provide a natural reference for interpreting metrics. On both French FDBs, humans show low pause-handling TOR and strong turn-taking and interruption TOR. However, they have higher interruption latency than non-cascaded English models, suggesting that natural dialogue is not only about responding quickly, but also about respecting pauses and taking appropriate turns. On MEDIA-FDB, the French cascade receives a higher rating than humans, showing that quality and naturalness may differ. Thus, optimizing FDB metrics does not always align with conversational naturalness.

\section{Conclusion}
We introduced CALLFC-FDB and MEDIA-FDB, two French Full-Duplex-Bench datasets derived from real-world conversations in Canadian and European French. These benchmarks provide a practical evaluation resource for French SDMs while remaining compatible with the FDB framework. Our results show that evaluation language affects full-duplex skills differently. Timing-related behaviors, such as pause handling TOR and interruption latency, remain relatively stable across languages and are largely driven by system architecture. In contrast, content-dependent behaviors, particularly user interruption rating, degrade when the model and benchmark language do not match. In contrast, content-dependent behaviors, particularly user interruption rating, degrade when the model and benchmark language do not match. In practice, evaluation language, system architecture, and latency should be considered together when selecting and evaluating full-duplex systems.

We also find that human--human conversations do not always yield optimal FDB scores, suggesting that benchmark performance alone may not fully reflect human-like interaction. Overall, our findings highlight the need to distinguish between acoustic-timing competence and language-dependent interaction competence in multilingual full-duplex evaluation. Future work will investigate whether this language dependency extends to other FDB v1.5 and v2.0 tasks and will expand the French FDBs to overlapping speech scenarios.

{
\makeatletter
\ifacl@finalcopy
\section*{Acknowledgments}
We acknowledge the support of the Natural Sciences and Engineering Research Council of Canada (NSERC) for this work, 
and would also like to thank Ministry of Economy and Innovation (MEI) of the Government of Québec for its continued support. 
\fi
}

\section*{Limitations}

This work adapts the French Full-Duplex-Bench only for FDB v1 tasks. Therefore, future work should extend the proposed French benchmarks to FDB v1.5 and FDB v2, including overlapping speech and multi-turn scenarios.

The latency comparison is also limited by the lack of a fast French speech-to-speech model. More comparable French and multilingual systems are needed to better isolate the effect of language from system architecture.

The language comparison varies in corpus, domain, and conversational formality, since the English FDB, CALLFC-FDB, and MEDIA-FDB are derived from different interaction settings. Future work with closely matched multilingual datasets could better isolate the effect of language from these corpus-level differences.

This study only evaluates language mismatch for FDB v1. It remains important to examine how mismatched evaluation languages affect FDB v1.5 and FDB v2 tasks.

Finally, FDB metrics mainly evaluate timing and turn-taking, but they do not fully capture conversational naturalness, user satisfaction, or long-term dialogue coherence. Benchmark scores should therefore be interpreted together with qualitative analysis and human-centered evaluation.

\section*{Ethics Statement}

We release only annotations and preparation scripts, not original or extracted audio. Users must obtain CALLFRIEND and MEDIA from their official distributors and follow the corresponding licenses. Samples are generated locally and remain subject to the original dataset licenses. We also do not release speaker metadata or personal identifiers, limiting additional privacy risks.

\bibliography{latex/luqia-dialoral}

\appendix

\section{Confidence Intervals}
\label{sec:confidence_intervals}
Table~\ref{tab:all_confidence_intervals} reports the 95\% confidence
intervals for all results in Table~\ref{tab:fdb_tasks_EN_FR}. 

\begin{table*}[!t]
    \centering
    \small
    \setlength{\tabcolsep}{4pt}
    \resizebox{\textwidth}{!}{
        \begin{tabular}{llccccccccc}
            \toprule
            \textbf{Data}
            & \textbf{Model}
            & \multicolumn{1}{c}{\textbf{Pause Handling}}
            & \multicolumn{3}{c}{\textbf{Backchannel}}
            & \multicolumn{2}{c}{\textbf{Smooth Turn Taking}}
            & \multicolumn{3}{c}{\textbf{User Interruption}} \\
            \cmidrule(lr){3-3}
            \cmidrule(lr){4-6}
            \cmidrule(lr){7-8}
            \cmidrule(lr){9-11}
            &
            & TOR $\downarrow$
            & TOR $\downarrow$
            & Freq $\uparrow$
            & JSD $\downarrow$
            & TOR $\uparrow$
            & Latency $\downarrow$
            & TOR $\uparrow$
            & Rating $\uparrow$
            & Latency $\downarrow$ \\
            \midrule

            \multirow{4}{*}{English FDB}
            & Freeze-Omni (En) & [0.4776, 0.5819] & [0.5260, 0.7831] & [0.0137, 0.0481] & [0.8437, 0.9298] & [0.2347, 0.4040] & [0.7752, 1.0058] & [0.4959, 0.6341] & [3.0150, 3.6930] & [0.8443, 1.2909] \\
            & PersonaPlex (En) & [0.2524, 0.3482] & [0.1368, 0.3723] & [0.0788, 0.1128] & [0.7410, 0.7716] & [0.8985, 0.9839] & [0.2105, 0.4329] & [0.9261, 0.9839] & [4.5794, 4.8238] & [0.1300, 0.1641] \\
            & Cascade (En) & [-0.0011, 0.0181] & [0.0132, 0.1686] & [-0.0006, 0.0048] & [0.9764, 1.0015] & [0.4388, 0.6200] & [6.5450, 7.7680] & [0.3123, 0.4477] & [2.8374, 3.5836] & [1.9345, 3.2886] \\
            & Cascade (Fr) & [-0.0022, 0.0135] & [0.0776, 0.2861] & [0.0023, 0.0105] & [0.9353, 0.9847] & [0.1216, 0.2650] & [7.0251, 8.6975] & [0.1577, 0.2723] & [-0.0330, 0.1725] & [3.2743, 6.1360] \\

            \midrule
            \multirow{5}{*}{CALLFC-FDB}
            & Freeze-Omni (En) & [0.46, 0.62] & [0.10, 0.39] & [0.00, 0.02] & [0.92, 0.98] & [0.28, 0.44] & [0.99, 1.21] & [0.51, 0.67] & [0.19, 0.50] & [0.75, 1.46] \\
            & PersonaPlex (En) & [0.37, 0.53] & [0.32, 0.65] & [0.08, 0.14] & [0.78, 0.85] & [0.72, 0.85] & [0.74, 1.02] & [0.66, 0.80] & [0.04, 0.25] & [0.28, 0.42] \\
            & Cascade (En) & [0.15, 0.28] & [0.49, 0.81] & [0.00, 0.01] & [0.95, 0.99] & [0.09, 0.20] & [7.06, 8.58] & [0.32, 0.48] & [0.51, 1.16] & [3.98, 6.62] \\
            & Cascade (Fr) & [0.07, 0.18] & [0.32, 0.65] & [0.01, 0.03] & [0.88, 0.95] & [0.23, 0.38] & [7.62, 8.88] & [0.92, 0.99] & [1.14, 1.63] & [4.45, 6.01] \\
            & Human-Human (Fr) & [0.13, 0.26] & [0.12, 0.42] & [0.09, 0.14] & [0.54, 0.64] & [0.71, 0.84] & [0.09, 0.25] & [0.80, 0.92] & [1.70, 2.16] & [1.12, 2.05] \\

            \midrule
            \multirow{5}{*}{MEDIA-FDB}
            & Freeze-Omni (En) & [0.46, 0.62] & -- & -- & -- & [0.31, 0.51] & [0.95, 1.14] & [0.47, 0.79] & [-0.05, 0.89] & [0.46, 1.58] \\
            & PersonaPlex (En) & [0.16, 0.29] & -- & -- & -- & [0.72, 0.88] & [0.97, 1.29] & [0.76, 0.98] & [-0.06, 0.37] & [0.19, 1.01] \\
            & Cascade (En) & [0.02, 0.10] & -- & -- & -- & [0.40, 0.60] & [7.67, 8.56] & [0.92, 1.03] & [0.15, 0.82] & [2.60, 5.54] \\
            & Cascade (Fr) & [0.12, 0.24] & -- & -- & -- & [0.34, 0.54] & [7.66, 8.70] & [0.79, 1.00] & [3.06, 3.88] & [3.07, 6.08] \\
            & Human-Human (Fr) & [0.01, 0.07] & -- & -- & -- & [0.97, 1.01] & [0.29, 0.37] & [0.83, 1.01] & [2.23, 3.42] & [1.13, 3.18] \\

            \bottomrule
        \end{tabular}
    }

\caption{95\% confidence intervals for all results reported in
Table~\ref{tab:fdb_tasks_EN_FR}.}
\label{tab:all_confidence_intervals}
\end{table*}
\end{document}